\documentstyle[12pt]{article}
\catcode`\@=11

\def\@normalsize{\@setsize\normalsize{15pt}\xiipt\@xiipt
\abovedisplayskip 14pt plus3pt minus3pt%
\belowdisplayskip \abovedisplayskip
\abovedisplayshortskip  \z@ plus3pt%
\belowdisplayshortskip  7pt plus3.5pt minus0pt}
\def\small{\@setsize\small{13.6pt}\xipt\@xipt
\abovedisplayskip 13pt plus3pt minus3pt%
\belowdisplayskip \abovedisplayskip
\abovedisplayshortskip  \z@ plus3pt%
\belowdisplayshortskip  7pt plus3.5pt minus0pt
\def\@listi{\parsep 4.5pt plus 2pt minus 1pt
            \itemsep \parsep
            \topsep 9pt plus 3pt minus 3pt}}

\def\underline#1{\relax\ifmmode\@@underline#1\else
        $\@@underline{\hbox{#1}}$\relax\fi}
\@twosidetrue
\relax

\catcode`@=12

\catcode`\@=11

\def\section{\@startsection{section}{1}{\z@}{3.5ex plus 1ex minus
   .2ex}{2.3ex plus .2ex}{\large\bf}}

\def\ps@headings{\def\@oddfoot{}\def\@evenfoot{}
\def\@oddhead{\hbox{}\hfill
        \makebox[.5\textwidth]{\raggedright\ignorespaces --\thepage{}--
        \hfill }}
\def\@evenhead{\@oddhead}
\def\subsectionmark##1{\markboth{##1}{}}
}

\ps@headings

\catcode`\@=12

\relax

\def\figcap{\section*{Figure Captions\markboth
        {FIGURECAPTIONS}{FIGURECAPTIONS}}\list
        {Fig. \arabic{enumi}:\hfill}{\settowidth\labelwidth{Fig. 999:}
        \leftmargin\labelwidth
        \advance\leftmargin\labelsep\usecounter{enumi}}}
 \relax
\def\tablecap{\section*{Table Captions\markboth
        {TABLECAPTIONS}{TABLECAPTIONS}}\list
        {Table \arabic{enumi}:\hfill}{\settowidth\labelwidth{Table 999:}
        \leftmargin\labelwidth
        \advance\leftmargin\labelsep\usecounter{enumi}}}
 \relax
\def\reflist{\section*{References\markboth
        {REFLIST}{REFLIST}}\list
        {[\arabic{enumi}]\hfill}{\settowidth\labelwidth{[999]}
        \leftmargin\labelwidth
        \advance\leftmargin\labelsep\usecounter{enumi}}}
 \relax

\catcode`\@=11

\def\marginnote#1{}
\newcount\hour
\newcount\minute
\newtoks\amorpm
\hour=\time\divide\hour by60
\minute=\time{\multiply\hour by60 \global\advance\minute by-
\hour}
\edef\standardtime{{\ifnum\hour<12 \global\amorpm={am}%
    \else\global\amorpm={pm}\advance\hour by-12 \fi
    \ifnum\hour=0 \hour=12 \fi
    \number\hour:\ifnum\minute<100\fi\number\minute\the\amorpm}}
\edef\militarytime{\number\hour:\ifnum\minute<100\fi\number\minute}
\def\draftlabel#1{{\@bsphack\if@filesw {\let\thepage\relax
  \xdef\@gtempa{\write\@auxout{\string
    \newlabel{#1}{{\@currentlabel}{\thepage}}}}}\@gtempa
    \if@nobreak \ifvmode\nobreak\fi\fi\fi\@esphack}
     \gdef\@eqnlabel{#1}}
\def\@eqnlabel{}
\def\@vacuum{}
\def\draftmarginnote#1{\marginpar{\raggedright\scriptsize\tt#1}}
\def\draft{\oddsidemargin -.5truein
        \def\@oddfoot{\sl preliminary draft \hfil
        \rm\thepage\hfil\sl\today\quad\militarytime}
        \let\@evenfoot\@oddfoot \overfullrule 3pt
        \let\label=\draftlabel
        \let\marginnote=\draftmarginnote
   
\def\@eqnnum{(\theequation)\rlap{\kern\marginparsep\tt\@eqnlabel}%
\global\let\@eqnlabel\@vacuum}  }
\def\preprint{\twocolumn\sloppy\flushbottom\parindent 1em
        \leftmargini 2em\leftmarginv .5em\leftmarginvi .5em
        \oddsidemargin -.5in    \evensidemargin -.5in
        \columnsep 15mm \footheight 0pt
        \textwidth 250mmin      \topmargin  -.4in
        \headheight 12pt \topskip .4in
        \textheight 175mm
        \footskip 0pt
        
\def\@oddhead{\thepage\hfil\addtocounter{page}{1}\thepage}
        \let\@evenhead\@oddhead \def\@oddfoot{} \def\@evenfoot{} 
}
\def\titlepage{\@restonecolfalse\if@twocolumn\@restonecoltrue\onecolumn
     \else \newpage \fi \thispagestyle{empty}\c@page\z@
        \def\thefootnote{\fnsymbol{footnote}} }
\def\endtitlepage{\if@restonecol\twocolumn \else  \fi
        \def\thefootnote{\arabic{footnote}}
        \setcounter{footnote}{0}}  
\catcode`@=12
\relax

\def\ps@headings{\def\@oddfoot{}\def\@evenfoot{}
\def\@oddhead{\hbox{}\hfill
        \makebox[.5\textwidth]{\raggedright\ignorespaces --\thepage{}--
        \hfill }}
\def\@evenhead{\@oddhead}
\def\subsectionmark##1{\markboth{##1}{}}
}

\ps@headings

\relax

\def\firstpage#1#2#3#4#5#6{
\begin{document}
\begin{titlepage}
\nopagebreak
\title{\begin{flushright}
        \vspace*{-1.8in}
        {\normalsize CERN-TH/98-79}\\[-9mm]
        {\normalsize hep-th/9803109}\\[4mm]
\end{flushright}
\vspace{2.5cm}
{#3}}
\author{\large #4 \\[0.0cm] #5}
\maketitle
\vskip 3mm
\nopagebreak 
\begin{abstract}
{\noindent #6}
\end{abstract}
\vfill
\begin{flushleft}
\rule{16.1cm}{0.2mm}\\[-3mm]
{\small$^\dag$ e-mail: Sergio.Ferrara, Alexandros.Kehagias, 
Herve.Partouche, Alberto.Zaffaroni@cern.ch} \\[-3mm]
CERN-TH-98-79\\[-3mm]
March 1998
\end{flushleft}
\thispagestyle{empty}
\end{titlepage}}

\def\simlt{\stackrel{<}{{}_\sim}}
\def\simgt{\stackrel{>}{{}_\sim}}
\newcommand{\dal}{\raisebox{0.085cm}
{\fbox{\rule{0cm}{0.07cm}\,}}}

\newcommand{\be}{\begin{eqnarray}}
\newcommand{\ee}{\end{eqnarray}}
\newcommand{\btau}{\bar{\tau}}
\newcommand{\p}{\partial}
\newcommand{\bp}{\bar{\partial}}
\newcommand{\cR}{{\cal R}}
\newcommand{\tR}{\tilde{R}}
\newcommand{\tcR}{\tilde{\cal R}}
\newcommand{\hR}{\hat{R}}
\newcommand{\hcR}{\hat{\cal R}}
\newcommand{\oE}{\stackrel{\circ}{E}}
\renewcommand{\p}{\partial}
\renewcommand{\bp}{\bar{\partial}}

\newcommand{\gsi}{\,\raisebox{-0.13cm}{$\stackrel{\textstyle
>}{\textstyle\sim}$}\,}
\newcommand{\lsi}{\,\raisebox{-0.13cm}{$\stackrel{\textstyle
<}{\textstyle\sim}$}\,}
\date{}
\firstpage{3118}{IC/95/34}
{\large {\Large M}EMBRANES AND {\Large F}IVEBRANES WITH {\Large L}OWER \\
{\Large S}UPERSYMMETRY   
AND {\Large T}HEIR {\Large A}d{\Large S}   
{\Large S}UPERGRAVITY  {\Large D}UALS \\
\phantom{X}}
{S. Ferrara, A. Kehagias,  H. Partouche and A. Zaffaroni$^\dag$} 
{
\normalsize\sl Theory Division, CERN, 1211 Geneva 23, Switzerland
}
{We consider superconformal field theories in three and six dimensions with
eight supercharges which can be realized on the world-volume of  M-theory 
branes sitting at orbifold singularities. We find that they should admit a 
${\cal N}=4$ and ${\cal N}=2$ supergravity dual in 
$AdS_4$ and $AdS_7$, respectively. 
We discuss the characteristics of the corresponding gauged supergravities.
}
In recent papers, the close connection between $AdS_{p+2}$ and the dynamics
on the world-volume of p-branes was explored \cite{kleb,sken,skentwo,kall,gun,torin}. The explicit  
proposal of Maldacena \cite{malda} that the large $N$ limit of certain conformal field 
theories can be described in terms of supergravity paved the way for 
a novel approach to superconformal theories. In particular, it has been argued that 
the type IIB supergravity on $AdS_5\times S^5$ is dual to $D=3$, 
${\cal N}=4$ 
$U(N)$ SYM theory at large $N$.  
Similarly, it has been proposed that the $(2,0)$, $D=6$ 
superconformal field theory with $N$ tensor multiplets 
is dual for large $N$  to eleven-dimensional supergravity
on $AdS_7\times S^4$ with large radii. It has also  been conjectured that
these dualities can be elevated to field-theory/string-theory 
and field-theory/M-theory equivalence for finite $N$.

The proposed duality between large $N$ field theories and anti-de Sitter
supergravities has been tested by identifying massless excitations in 
the bulk with singletons composite operators on the anti-de Sitter boundary \cite{fer}.
Extending further this relation, it has been argued that in a suitable limit,
the generating functional for the boundary correlators of singleton
composite field is reproduced by the anti-de Sitter supergravity action 
\cite{pol,witten}. The conjecture was further explored in 
\cite{ooguri,IMSY,FFZ,sjr,maldathree,arefa,behr,lu}.
Moreover, models with lower world-volume supersymmetries started to be 
explored \cite{KS,berkooz,LNV,BKV}. We could ask  if the  field-theory/string-theory 
correspondence can be extended in such a way that  for any given ${\cal N}=0,1,2$ superconformal 
model in four dimensions there exist a supergravity theory in 
$AdS_5$, and, similarly, 
if the field-theory/M-theory duality  gives a supergravity
theory on $AdS_4$ or $AdS_7$ for any superconformal theory in three and six 
dimensions, respectively. The conjecture, in the form in which it has been 
formulated up to now, requires the realization of the superconformal theory in terms of branes, in a setting which is compatible with the limits considered in \cite{malda}. Fortunately enough, quite a lot of superconformal models can be realized in this way. It is one of the purpose of this paper, to show superconformal
models in $D=3,6$ which fulfil the previous requirements.

The maximally supersymmetric theories
in three and six dimensions have been explored recently \cite{aha,LR,min,H}. 
Here, we will consider $D=3$ and $D=6$ theories with lower supersymmetries, 
in particular, ${\cal N}=4$ and ${\cal N}=1$ supersymmetric 
theories realized on the world-volume of  M2- and M5-branes. 
Such theories can be obtained by appropriate orbifolds of the transverse
space. For the M2-brane, the transverse space is ${\bf C}^4$ and one may 
consider orbifolds of the form $ 
{\bf C}^2/\Gamma\times
{\bf C}^2/\Gamma',~~{\bf C}\times {\bf C}^3/\Gamma, or {\bf C}^4/\Gamma,$ 
corresponding to M-theory compactifications on $K3\times K3$, $CY_3$
and $CY_4$, respectively, giving rise to theories with ${\cal N}=4$,
${\cal N}=2$ and ${\cal N}=1$. For the M5-brane, the transverse space 
is ${\bf R}^5$ and one
may consider orbifolds ${\bf C}^2/\Gamma\times {\bf R}$ that  
 lead to 
$(1,0)$ theories on the world-volume of the M5-brane. A similar procedure was
followed in \cite{KS} and further elaborated in \cite{LNV} for theories
on the D3-brane of the type IIB string theory.

Let us start with $D=6$.
The  M5-brane  breaks half of the 32 supersymmetries leaving a 
$(2,0)$ theory on the M5 world-volume. There exist   
 five scalar fields  representing fluctuations transverse to the 
brane and   three additional degrees of freedom      
coming as collective coordinates 
associated to the three-form of the eleven-dimensional 
supergravity. 
All together, we have 8 degrees of freedom which 
fill the unique ${\cal N}=2$, $D=6$
tensor multiplet of the chiral $(2,0)$, $D=6$ supersymmetry. The 
eleven-dimensional supergravity solution for $N$  M5-branes is 
\be
ds^2&=&f^{-1/3}\left(-dt^2+dx^adx^a\right)+f^{2/3}dy^ady^a\, ,
\nonumber \\
f&=& 1+\frac{\pi N\ell_p^3}{r^3}\, , ~~~r^2=y^ay^a\, , 
\ee
where $\ell_p$ is the eleven-dimensional Planck length, $x^a,y^a$ 
$(a=1,..,5)$ 
are  coordinates on the five-brane and transverse to it, 
respectively. There 
is a horizon at $r=0$. The near horizon geometry is given by $AdS_7\times S^4$,
namely as the product of a seven-dimensional anti-de Sitter space with a
four-sphere of radii $R_{S^4}=R_{AdS_7}/2=\ell_p(\pi N)^{1/3}$. The M-theory 
five-brane interpolates between flat Minkowski space-time at $r\to\infty$
and $AdS_7\times S^4$ at the horizon. The 
decoupling limit 
considered in \cite{malda} is $\ell_p\to 0$ and  $r/{\ell_p}^3$  finite, 
while  supergravity is valid for $N>>1$. 

In order 
to describe a gauge theory in six dimensions, we should further break the 
$(2,0)$ supersymmetry to the  $(1,0)$ one. The massless sector of the latter
contains  tensors, vectors and 
 hypermultiplets. In particular, 
the $(2,0)$ tensor multiplet gives rise to a  tensor 
and a hyper of the $(1,0)$ superalgebra. A $(1,0)$ supersymmetry 
on the M5-brane can be obtained by taking an appropriate 
orbifold in the transverse space. 
Since we want to keep in the near-horizon geometry the $AdS_7$ structure, the
orbifolding should be such that it acts on $S^4$ only. Now at 
$t,x^a=const.$, the transverse space is topologically ${\bf R}^5$ with metric
\be
 ds_\perp=\left(1+\frac{\pi N\ell_p^3}{r^3}\right)^{2/3} dy^ady^a\, . 
\ee
We may act with a discrete group $\Gamma\subset SU(2)$ on ${\bf R}^5$ 
to form 
${\bf R}^5/\Gamma={\bf C}^2/\Gamma\times {\bf R}$ where the modding by 
$\Gamma$ identifies points which are at equal distance from the origin. 
Thus, the transverse 
space to the M5-branes  is the 
product of a flat sixth direction ${\bf R}$ and 
the ALE space ${\bf C}^2/\Gamma$. We will consider here the cases 
$\Gamma={\bf Z}_k$ $(k\geq2)$ and $\Gamma={\cal D}_{k}$ $(k\geq 4)$ where  
${\cal D}_{k}$ is the binary extension of the dihedral group.
The former gives an
$A_{k-1}$ ALE space and the latter a $D_k$ one.
The generators of $\Gamma$  act 
on ${\bf C}^2$ with coordinates $z^1,z^2$ as
\be
{\bf Z}_k~:&& ~~z^1\to e^{2i\pi\over k}z^1\, , ~~~
z^2\to e^{-{2i\pi\over k}}z^2\, ,  \\
{\cal D}_k~:&& ~~~~z^1\to e^{i\pi/(k-2)}z^1\, , ~~~
z^2\to e^{-{2i\pi/(k-2)}}z^2~~~ \mbox{and}~~~z^1\to z^2 ,~~~
 z^2\to -z_1\, .  \nonumber 
\ee
Since these ALE  spaces are of  
$SU(2)$ holonomy, they break half of the original 32 
supersymmetries while half more are broken by  the M5-branes. Thus, the   
eight remaining supercharges   
lead to a $(1,0)$
world-volume theory in six dimensions. 
The $(2,0)$ tensor multiplet gives a $(1,0)$ tensor 
and a  hyper. The tensor contains an  antiself-dual two-form and
a scalar representing fluctuations of the five-brane along the sixth  direction
${\bf R}$, which is not affected by the orbifold. 
The hyper contains two complex scalars $Z^1,Z^2$ 
corresponding to the position of the M5-branes  
in the ALE space. In addition, there 
exist vector multiplets coming from the wrapping of M-theory membranes along
the non-vanishing two-cycles of the ALE space. 

We will first  consider the case of $N$ M5-branes with world-volume 
$(0,1,2,3,4,5)$ located at $N$ points in 
the sixth direction ${\bf R}$ and at the same point in the $A_{k-1}$ 
ALE space. 
To analyze the spectrum, it is more convenient to consider our system
in the context of type IIA theory. Let us recall that a metric of the form
\be
ds_{TN}^2&=&V^{-1}(dX_{11}+\vec{\omega})+Vd\vec{X}d\vec{X}\, , \nonumber \\
V&=&1+\sum_{i=1}^k\frac{R_{11}}{2|\vec{X}-\vec{X}_i|}\, , ~~~\vec{\nabla}
\times \vec{\omega}=-\vec{\nabla}V\, , \label{NUT}
\ee
where $\vec{X}$ is a three vector  and 
$X_{11}$ has periodicity $2\pi R_{11}$ describes a Taub-NUT space.
In the limit $R_{11}\to \infty$ we can ignore the $1$ in the expression of 
$V$ and the Taub-NUT space approaches the ALE 
space ${\bf C}^2/{\bf Z}_k$. By identifying $R_{11}$ with the 
type IIA string coupling $g^A=(R_{11}/\ell_p)^{3/2}$, 
we may  replace the ALE space in M-theory  with a Taub-NUT one.  
This can be interpreted as M-theory with $k$ KK monopoles, which
 is type IIA with $k$ D6-branes
located at $ \vec{X}_i$ \cite{firstseib,pz,WI,sen}. Thus, the spectrum 
of  $N$ M5-branes with transverse space ${\bf R}\times {\bf C}^2/{\bf Z}_k$
in M-theory for large  $R_{11}$ can be analyzed in the  type IIA side by
considering  $k$ D6-branes and $N$ NS5-branes at strong string coupling. 
One should expect that the spectrum can be evaluated at weak 
coupling and followed into strong coupling by rescaling appropriate 
parameters and moduli.

The spectrum of the above system at weak coupling can be analyzed by standard
brane technology \cite{karch,hz,karchtwo}. 
We   consider a system of $N$ NS5-branes with world-volume $(0,1,2,3,4,5)$
and $k$ D6-branes stretched between them with world-volume $(0,1,2,3,4,5,6)$.
This configuration  breaks the ten-dimensional 
Lorentz-invariance to $SO(1,5)\times
SO(3)$ and has a $(1,0)$, D=6 supersymmetry living on the world-volume of 
the NS5-branes. The latter has an $Sp(1)$ R-symmetry which is geometrically
realized as rotations in the three-dimensional space transverse to the 
D6-brane (the $SO(3)$ group above.) We choose the $N$ NS5-branes to be 
located at $(x^6_\alpha,x^7,x^8,x^9)$, $\alpha=1,...,N$,  
and coincident with $k$ D6-branes at the same fixed coordinate $(x^7,x^8,x^9)$
in the transverse space. If the NS5-branes were absent, we 
would have a system of $k$ coincident D6-branes,  which in the M-theory 
language would be interpreted as $k$  KK-monopoles. In this 
case, there would be an $U(k)$ gauge symmetry. The 
presence of the NS5-branes implies that the D6-branes are in fact divided
in $N-1$ pieces of finite length along the sixth direction, each of them 
stretched between two adjacent NS5-branes. In addition, there are
$k$ semi-infinite D6-branes on the left and right of the first and last 
NS5-branes, respectively.    
This  has the effect to give rise to the 
gauge group $U(k)^{N-1}$ on the NS5-branes world-volume as follows from the
KK reduction of the D6 world-volume theory along $x^6$.   There
exist charged hypermultiplets in the representation 
\be
&&({\bf k},{\bf \bar{k}},1,...,1)\oplus (1,{\bf k},{\bf \bar{k}},1,...,1)\oplus
\cdots \oplus(1,...,1,{\bf k},{\bf \bar{k}})\oplus
\nonumber\\
&& k({\bf \bar{k}},1,...,1)\oplus k(1,...,1,{\bf k})
\ee
of the gauge group, where the two last terms arise from the semi-infinite 
D6-branes. In addition to the massless spectrum comming from the 
D6-branes, each  of the NS5-branes provides a $(2,0)$ tensor 
multiplet.
In addition to the self-dual two-forms, these $N$ tensor multiplets of the 
$(2,0)$ 
superalgebra contain five scalars 
$\phi^I_\alpha$ $(I=6,...,10)$. Under the unbroken $(1,0)$ supersymmetry, 
these multiplets  decompose
into  tensor multiplets that contain the scalars $\phi^6_\alpha$, while 
$\phi_\alpha^{7,8,9,10}$ fill   hypermultiplets. 
The KK reduction to $(5+1)$ dimensions shows that $\phi_\alpha^6$
appears as the effective coupling of each of the $N-1$ $U(k)$ factors
\be
\frac{1}{g_\alpha^2}{F^\alpha_{\mu\nu}}^2+
(\partial\phi^6_\alpha)^2+\sqrt{c}\phi_\alpha^6 {F^\alpha_{\mu\nu}}^2\, .
\label{gg}
\ee
The bare coupling $g_\alpha$ can be absorbed into $\phi_\alpha^6$ so that 
the effective coupling is indeed 
\be
\frac{1}{{g_\alpha^{eff}}^2}=\sqrt{c}\phi^6_{\alpha}\, , 
\label{geff}
\ee
 where $c$ is the anomaly
coefficient \cite{Ssix}. 
$\phi^{7,8,9}_\alpha$ are  $SO(3)$-triplets  of FI terms 
for the diagonal $U(1)$'s of the gauge group factors. 
The hypermultiplets they belong to  are exactly what is needed 
to cancel the $U(1)$ anomalies and, as a consequense of a Green-Schwarz 
mechanism,  all the diagonal $U(1)$ gauge bosons are massive. 
As a result, 
the $(5+1)$ world-volume theory we obtain is the ${\cal N}=1$, $SU(k)^{N-1}$ 
gauge theory with one tensor multiplet and $k$ hypermultiplets in the 
${\bf k}\oplus
{\bf \bar{ k}}$ representation for each of the gauge group factors.

Six-dimensional gauge theories are restricted by the gauge anomaly, 
which  is given by  
\be
Tr_{adj}{F}^4-Tr_{R}{F}^4
=a\;tr{F}^4+
\frac{c}{d^2}(tr{F}^2)^2 \label{anomaly}
\ee
for a simple group, 
where $Tr_{adj},Tr_{R}$ and $tr$ 
are traces in the adjoint, 
$R$ and fundamental 
representations, 
respectively, and $d$ is the dimension of the fundamenantal. 
When the theory
is not coupled to gravity, it makes sence  when $\alpha=0$ and $c\geq0$. 
For $c=0$, the theory 
is anomaly free, while for $c>0$
the anomaly can be canceled by a tensor multiplet \cite{sagnotti,Ssix}. 
In our case, 
$c=3k^2$ for each $SU(k)$ factor and the anomaly can indeed be canceled by 
the tensor multiplet.  
In fact, as follows from eq.(\ref{geff}), 
the theory may have a non-trivial fixed point at $\phi^6_\alpha=0$ where it  
is strongly coupled. After absorbing the bare coupling into 
$\phi^6_\alpha$, 
the terms in (\ref{gg}) are scale invariant, supporting 
the existence of this non-trivial fixed point. 
In fact, in the large $N$ limit, the six-dimensional 
 $(1,0)$ $SU(k)^{N-1}$ 
SYM theory with one tensor and $k$ hypermultiplets in 
${\bf k}\oplus{\bf \bar{k}}$ for each 
$SU(k)$ factor  should be dual to 
eleven-dimensional supergravity on $AdS_7\times S^4/{\bf Z}_k$ of radii
$R_{S^4}=R_{AdS_7}/2=\ell_p(\pi N)^{1/3}$ and therefore
conformal invariant. Then, it is natural to conjure that for all $N$,
the above six-dimensional $(1,0)$ $SU(k)^{N-1}$ 
SYM theory is  M-theory on 
$AdS_7\times S^4/{\bf Z}_k$.

We will now consider the case where $\Gamma$ is the binary dihedral 
group which corresponds to a $D_k$ ALE space.  
It is known that M-theory on such spaces is type IIA with  
$2k$ D6-branes  sitting at  an orientifold O6-plane with charge $-4$. 
Therefore, the system we consider is as in the  $A_{k-1}$ case, namely 
$N$ NS5-branes with world-volume $(0,1,2,3,4,5)$ and D6-branes stretched 
between them, with world-volume $(0,1,2,3,4,5,6)$, together with an 
orientifold six-plane. The system is located at a single point in the 
transverse space. Charge conservation imposes $N$ to be even and the 
gauge group turns out to be \cite{karch,hz,karchtwo} $Sp(k-4)\times 
\Big{(}SO(2k)\times Sp(k-4)\Big{)}^{N/2-1}$. 
Each factor of the gauge group is coupled to a $(1,0)$ 
tensor multiplet as in eq.(\ref{gg}) and the 
hypermultiplet content is
\be
&&\frac{1}{2}\Big{\{}\left({\bf 2k-8},{\bf 2k},1,...,1\right)\oplus 
\left(1,{\bf 2k},{\bf 2k-8},1,...,1\right)\oplus\cdots
\oplus
 \left(1,...,1,{\bf 2k-8},{\bf 2k},1\right)
\oplus \nonumber \\&&
\left(1,...,1,{\bf 2k},{\bf 2k-8}\right)\oplus 
2k\left({\bf 2k-8},1,...,1\right)\oplus 2k\left(1,...,1,{\bf 2k-8}\right)
\Big{\}}
\, .
\ee 
Effectively, each $SO(2k)$ factor of the gauge group is coupled to $2k-8$ 
hypers in the vectorial representation ${\bf 2k}$, while each of the 
$Sp(k-4)$ factor is coupled to $2k$ hypers in the fundamental ${\bf 2k-8}$. 
In this case the anomaly eq.(\ref{anomaly}) has $a=0$ and $c=12k^2$,
$c=12(k-4)^2$ for each $SO(2k)$ and $Sp(k-4)$ factor, respectively. 
Thus, the anomaly can be cancelled by the same mechanism as before, namely
by the coupling of the scalars in the tensor multiplets with the gauge fields. 
The theory flows then to a non-trivial fixed point in the IR which can be 
described by  the supergravity on $AdS_7\times S^5/{\cal D}_k$.  

Branes at orbifold singularities generally give rise to 
six-dimensional superconformal fixed points \cite{KI}. They  can be 
realized using
variations of the previous 
construction \cite{hz,karchtwo}. The ones which admit
a consistent description in M-theory (in some cases defined on $R/Z_2$--see
\cite{berkooz} for a related example) are likely to have a dual description
as a supergravity in $AdS_7$.

The ${\cal N}=2$ gauged supergravity, coupled to 
matter, on $AdS_7$ was constructed in \cite{menci}. The
gauge group is $SU(2)\times H$, where the $SU(2)$ vector 
fields, gauging the R-symmetry of the 
superconformal theory on the boundary, live in the supergraviton 
multiplet, while the vectors in the adjoint of $H$, gauging the 
flavour symmetries of the boundary theory, 
are provided by additional vector multiplets.
At the six-dimensional superconformal point discussed 
above (considering, for simplicity, 
the $A_k$ singularity), the global symmetry is $H=SU(k)$. 
The scalars in the supergravity theory parametrize the manifold,
\be
R^+\times {O(3,n)\over O(3)\times O(n)}\, ,
\ee
with $n=dim H$ and coordinates  a real scalar $\phi_0$ 
belonging to the graviton multiplet, and by the scalars 
$\phi_i^{\Lambda}$ in the adjoint of $SU(2)$ and Lie-algebra valued in $H$. 
The ${\cal N}=2$ supergravity has
a potential which admits a stable \cite{BF} $SU(2)$ invariant anti-de Sitter 
vacuum  at $\phi_0=\hat\phi_0, \phi_i^{\Lambda}=0$ \cite{menci}.

It would be interesting to understand better how the previous theory can
be obtained as a KK reduction of the eleven-dimensional supergravity to
seven dimensions \cite{van}, since the KK states would give information on the
spectrum of conformal operator of the superconformal theory living at the
boundary \cite{fer,witten,FFZ,aha,LR}. 
The relevant superconformal algebra is $OSp(6,2/2)$, which also
classifies the particle states in $AdS_7$.  Here we simply note that
singleton representations of the algebra, corresponding to degrees of freedom
living on the boundary of $AdS_7$, would correspond to the {\it fields} of
the superconformal theory, while the other representations correspond to 
composite operators \cite{fer}.  
  
Similar constructions can be repeated for the case of superconformal theories in three dimensions, by orbifolding the ${\cal N}=8$ 
example discussed in \cite{malda}.
The theory of $N$ coincident membranes in M theory is 
supposed to be dual for large $N$ to the eleven-dimensional 
supergravity on $AdS_4\times S^7$. We can
consider orbifolds of M-theory in which the $AdS_4$ part is not affected by the
projection. This corresponds to projecting by a suitable 
discrete group the transverse
space of the membranes $R^8$. 

${\cal N}=4$ superconformal theories in three dimensions have a $OSp(4/4)$
symmetry. Non-trivial superconformal theories can be obtained in the IR
limit of three dimensional Yang-Mills theories \cite{sw,si}. We will now determine a large class of three dimensional Yang-Mills theories, which have a
brane realization and whose IR limit should admit a dual description in terms
of a ${\cal N}=4$ supergravity in $AdS_4$.

The original ${\cal N}=8$ example in 
\cite{malda} can be also understood in the following way. 
Let us start with $N$ D2-branes in type IIA which give rise to 
the maximal supersymmetric Yang-Mills theory in three dimensions. 
The theory is not conformal, having a dimensionful gauge coupling 
but flows in the IR to a superconformal fixed point which is the 
same as the one discussed in \cite{malda}. The gauge coupling for 
the D2-branes is determined by the IIA string coupling in such a way 
that it becomes very large, and the theory therefore flows to the 
IR superconformal fixed point, exactly when the M-theory description
takes over and the system is better described with $N$ M-theory membranes. Note
also that the $SO(7)$ global symmetry of the D2-branes theory, which rotates
the transverse direction in type IIA, is promoted at the superconformal point
to an $SO(8)$ symmetry \cite{sethi}, which rotates the directions transverse 
to the membranes, and which is the appropriate 
R-symmetry of the ${\cal N}=8$ superconformal algebra.

Let us now consider the ${\cal N}=4$ case. 
The simplest example is obtained by putting
$N$ membranes near an orbifold singularity of the form 
$R^4/Z_k\times R^4/Z_n$.
The spectrum for this theory was computed in \cite{pz}. In a type IIA
description, we have a theory of D2-branes sitting at an orbifold 
singularity
in the presence of $n$ D6-branes, whose world-volume is, in part, 
parallel to the D2 world-volume and, for the remaining part, 
{\it wrapped} around a four-dimensional singular
space $R^4/Z_k$. The corresponding Yang-Mills theory realized on the 
world-volume of the D2-branes can be easily derived to be \cite{DM,pz} a 
$SU(N)^k$ theory
with hypermultiplets in the representations:
\be
({\bf N},{\bf N},1,...,1)\oplus (1,{\bf N},{\bf N},...,1)\oplus
\cdots\oplus({\bf N},1,...,1,{\bf N})+n({\bf N},1,...,1)\, .
\label{mirror}
\ee
In the IR, these theories flow to a superconformal fixed point 
which should have a
supergravity dual description as eleven-dimensional 
supergravity on $AdS_4\times S^7/(Z_k\times Z_n)$. 

Since there are two ways to get a type IIA description
starting from M theory on $R^4/Z_k\times R^4/Z_n$, which correspond to
{\it compactify}, in one case, on $R^4/Z_k$ and, 
in the second case,  on $R^4/Z_n$, we could expect to obtain two
Yang-Mills candidates (corresponding to the exchange of  $k$ and $n$) 
for the same IR fixed point (and the same supergravity description). However, and this was the original motivation of \cite{pz}, these two theories are actually three dimensional mirror pairs in 
the sense of \cite{si}, and therefore flow in the IR to the same superconformal fixed point.  The Yang-Mills theories described above indeed contain and generalize the example in \cite{si}.   
The differences between the theory in (\ref{mirror}) and its mirror, 
with $k$ and $n$
interchanged, in fact disappear in the eleven-dimensional description which is
the relevant one for capturing the structure of the superconformal fixed point.

The R-symmetry $SU(2)\times SU(2)$ of the superconformal point is already manifest in the Yang-Mills theory at finite coupling. In the ${\cal N}=4$ case, 
it is the
global flavour symmetry which is enhanced at the IR fixed point \cite{si}.
The manifest $SU(n)$ flavour symmetry of the theories discussed above
is enhanced in the IR to 
$SU(n)\times SU(k)$ 
(a symmetry which is manifest in the M-theory description). 

The case in which the orbifold projection is performed with a dihedral discrete
group is analogous. The Yang-Mills theories that can be obtained in this way
can be found in \cite{pz}.

These superconformal fixed points should have a description as a 
${\cal N}=4$ supergravity in $AdS_4$, whose supergraviton multiplet contains the $SU(2)\times SU(2)$ vector fields gauging the R-symmetry, coupled to additional vector multiplets which gauge the flavour symmetries. 
Such gauged supergravity was constructed
in \cite{roo}. If the gauge group is $SU(2)\times SU(2)\times H$, the scalars in the theory parametrize the manifold, 
\be
{SU(1,1)\over U(1)}\times {SO(6,n)\over SO(6)\times SO(n)}\, ,
\ee
with $n=dim H$. The coset $SU(1,1)/U(1)$ is parametrized by a
$SU(2)\times SU(2)$ singlet 
complex scalar $Z$ in the supergraviton multiplet. Since we are
looking for an anti-de Sitter  $SU(2)\times SU(2)$ invariant vacuum, all
the remaining scalars, in the adjoint of $SU(2)\times SU(2)$ and Lie-algebra 
valued in $H$, can be set to zero. 
The potential for $Z$ admits a stable $AdS_4$ vacuum 
when the coupling constants 
for the two $SU(2)$ are equal \cite{roo}, giving the
symmetric gauged $SO(4)$ supergravity. From the point of view of the
Yang-Mills theory which flows to this fixed point, the symmetry between the two
$SU(2)$ is nothing else than the mirror symmetry of \cite{si}.  

Also in these three-dimensional theories, it would be interesting to 
have a description in terms of an explicit 
KK reduction of the eleven-dimensional supergravity to
four dimensions, since the KK states would give information on the
spectrum of conformal operator of the superconformal theory living at the
boundary \cite{witten}. 
The relevant superconformal algebra is $OSp(4/4)$, which also
classifies the particle states in $AdS_4$. The irreducible representations of $OSp(4/4)$ are discussed in \cite{fronsdal}. The
singleton representations of the algebra, in this case,  correspond to a multiplet with a complex scalar and a fermion, transforming as $(1/2,0)+(0,1/2)$ under $SU(2)\times SU(2)$, in
the superconformal boundary theory, while the other representations correspond to 
composite operators \cite{fer}.

In principle, one could construct, by changing 
the orbifold projection, three-dimensional theories with 
lower ${\cal N}=2,1$ supersymmetries, based on the superconformal 
algebras $OSp(2/4)$ and $OSp(1/4)$. Their supergravity duals 
would correspond to 4d ${\cal N}=2,1$ $AdS_4$ supergravities with 
some additional matter multiplets.
The supermultiplets in $AdS_4$ would be composite operators of the singleton
representations on the 3d boundary \cite{fer}.

It would be interesting, both in three and six dimensions, to extend the 
results of the present paper to cases with lower supersymmetry and 
eventually to
cases with ${\cal N}=0$. This can be achieved by considering different types of
orbifold projections. In the dual supergravity side, it is known that the 
scalar potential for the gauged supergravities has other critical points where
supersymmetry is partially or completely broken.  We may ask if we can 
identify a superconformal boundary theory with lower or zero supersymmetry, 
whose generating functional for composite operators is reproduced
by the tree-level supergravity expanded around these other critical points. 
This would provide an explicit flow between superconformal theories with 
different supersymmetries.

\noindent{\bf Acknowledgements} 

This work is  supported in part by
the EEC under TMR contract ERBFMRX-CT96-0090.
S.F. is supported in part by the DOE under grant DE-FG03-91ER40662, 
Task C, and
by ECC Science Program SCI$^*$ -CI92-0789 (INFN-Frascati).


\begin{thebibliography}{99}
\bibitem{kleb} C. W. Gibbons and P. K. Townsend, Phys. Rev. Lett. 71 (1993) 3754; M. P. Blencowe and M. J. Duff, Phys. Lett. B203 (1988) 229; Nucl. Phys B310 (1988), 389; M. J. Duff, Class. Quantum Grav. 5 (1988) 189; E. Bergshoeff, M. J. Duff, C. N. Pope and E. Sezgin, Phys. Lett. B199 (1988) 69; H. Nicolai, E. Sezgin and Y. Tanii, Nucl. Phys B305 (1988) 483.
\bibitem{sken} K. Sfetsos and K. Skenderis, {\em ``Microscopic derivation of the Bekenstein-Hawking entropy formula for
     non-extremal black holes''}, hep-th/9711138.
\bibitem{skentwo} H. J. Boonstra, B. Peeters and K. Skenderis, {\em  ``Branes and Anti-de Sitter Space-Times''}, hep-th/9801076.
\bibitem{kall} P. Claus, R. Kallosh and A. Van Proeyen, 
{\em ``M5-brane and superconformal (0,2) tensor multiplet in 6 dimensions''}, hep-th/9711161;
R. Kallosh, J. Kumar and  A. Rajaraman, {\em ``Special Conformal Symmetry of World-volume Actions''}, hep-th/9712073;
P. Claus, R. Kallosh, J. Kumar, P. Townsend and A. Van Proeyen, {\em ``Conformal theory of M2, D3, M5 and D1+D5 branes''}, hep-th/9801206.
\bibitem{gun} M. Gunaydin and D. Minic, 
{\em ``Singletons, Doubletons and M-theory''}, hep-th/9802047.
\bibitem{torin} L. Castellani, A. Ceresole, R. D'Auria, S. Ferrara, P. Fr\'e and M. Trigiante, 
{\em ``G/H M-branes and $AdS_{p+2}$ Geometries}, hep-th/9803039.
\bibitem{malda} J. Maldacena, {\em ``The Large N Limit of Superconformal Field Theories and Supergravity''}, hep-th/9711200.
\bibitem{fer} S. Ferrara and C. Fronsdal, {\em ``Conformal 
Maxwell Theory as a Singleton Field Theory on $AdS_5$, IIB Three-Branes
     and Duality}, hep-th/971223; {\em ``Gauge Fields as Composite 
Boundary Excitations''}, hep-th/9802126.
\bibitem{pol} S. S. Gubser, I. R. Klebanov and A. M. Polyakov, {\em ``Gauge Theory Correlators from Non-Critical String Theory''}, hep-th/9802109.
\bibitem{witten} E. Witten, {\em ``Anti-de Sitter Space And Holography''},
hep-th/9802150.

\bibitem{ooguri} G. T. Horowitz and H. Ooguri, {\em ``Spectrum of Large N Gauge Theory from Supergravity''}, hep-th/9802116.
\bibitem{IMSY} N. Itzhaki, J.M. Maldacena,
J. Sonnenschein and  S. Yankielowicz, {\em ``Supergravity
and the Large N Limit of Theories with Sixteen Supercharges''},
hep-th/9802042.
\bibitem{FFZ} 
S. Ferrara, C. Fronsdal and A. Zaffaroni, {\em ``On N=8 Supergravity
on $AdS_5$ and N=4 Superconformal Yang-Mills Theory''}, hep-th/9802203;
S. Ferrara and A. Zaffaroni, {\em ``N=1, N=2 4-D Superconformal 
Field Theories and Supergravity in $AdS_5$''}, hep-th/9803060.

\bibitem{sjr} S. J. Rey and  J. Yee, {\em ``Macroscopic Strings as 
Heavy Quarks of Large N Gauge Theory and Anti-de Sitter
     Supergravity''}, hep-th/9803001.
\bibitem{maldathree} J. M. Maldacena, {\em ``Wilson loops in large 
N field theories'', hep-th/9803002.}
\bibitem{arefa} I. Ya. Aref'eva and I. V. Volovich, {\em ``Field 
Theories in Anti-de Sitter Space and Singletons''}, hep-th/9803028.
\bibitem{behr} E. Bergshoeff and  K. Behrndt, {\em ``D-Instantons 
and Asymptotic Geometries}, hep-th/9803090.
\bibitem{lu} M.J. Duff, H. Lu, C.N. Pope {\em ``Gauge Theory $\to$ 
IIB $\to$ IIA $\to$ M Duality''}, hep-th/9803061.
\bibitem{KS} S. Kachru and E. Silverstein, {\em ``4d Conformal Field Theories
and Strings on Orbifolds''}, hep-th/9802183.
\bibitem{berkooz} M. Berkooz, {\em ``A Supergravity Dual of a (1,0) 
Field Theory in Six Dimensions}, hep-th/9802195.
\bibitem{LNV}  A. Lawrence, N. Nekrasov and  C. Vafa, {\em ``On 
Conformal Field Theories in Four-Dimensions''}, 
   hep-th/9803015.


\bibitem{BKV} M. Bershadsky, Z. Kakushadze and  C. Vafa, 
{\em ``String Expansion as Large N Expansion of Gauge Theories''},
hep-th/9803076
\bibitem{aha} O. Aharony, Y. Oz and Z. Yin, {\em ``M-Theory 
on $AdS_p\times S^{11-p}$ and Superconformal Field Theories}, hep-th/9803051.

\bibitem{LR} R.G. Leigh and M. Rozali, {\em ``The Large N Limit of the $(2,0)$
Superconformal Field Theory''}, hep-th9803068.
\bibitem{min} S. Minwalla, {\em ``Particles on $AdS_{4/7}$ 
and Primary Operators on $M_{2/5}$ Brane World-volumes''}, hep-th/9803053.
\bibitem{H} E. Halyo, {\em ``Supergravity on $AdS_{4/7}\times
S^{7/4}$ and M Branes''}, hep-th/9803077. 


\bibitem{firstseib} N. Seiberg, Phys.Lett. B384 (1996) 81, hep-th/9606017.
\bibitem{pz} M. Porrati and A. Zaffaroni,  Nucl.Phys. B490 (1997) 107.
\bibitem{WI} E. Witten, Nucl.Phys. B500 (1997) 3, hep-th/9703166.
\bibitem{sen} A. Sen, {\em ``A Note on Enhanced Gauge Symmetries in M- and String Theory''}, hep-th/9707123.

\bibitem{karch} I. Brunner and A. Karch, Phys. 
Lett. B409 (1997) 109, hep-th/9705022.
\bibitem{hz} A. Hanany and A. Zaffaroni, {\em ``Branes and Six Dimensional Supersymmetric Theories}, hep-th/9712145.
\bibitem{karchtwo} I. Brunner and A. Karch, {\em ``Branes at Orbifolds versus Hanany Witten in Six Dimensions''}, hep-th/9712143.


\bibitem{Ssix} N. Seiberg, Phys. Lett. B390 (1997) 169, hep-th/9609161.
\bibitem{sagnotti} A. Sagnotti, Phys. Lett. B294 (1992) 196, hep-th/9210127.
\bibitem{KI} K. Intriligator, Nucl.Phys. B496 (1997) 17, 
hep-th/9702038; J. D. Blum and K. Intriligator, Nucl.Phys. B506 (1997) 22, hep-th/9705033; Nucl.Phys. B506 (1997) 
199, hep-th/9705044.
\bibitem{BF} P. Breitenlohner and D.Z. Freedman, Ann. Phys. 144 (1982) 197.
\bibitem{menci} L. Mencizescu, P. K. Townsend and P. van Nieuwenhuizen, Phys. Lett. 143B (1984) 584; E. Bergshoeff, I. G. Koh and E. Sezgin, Phys. Rev. D32
(1985) 1357.






\bibitem{van} P. van Nieuweinhuizen, Class. Quantum Grav. 2 (1985) 1.
\bibitem{sw} N. Seiberg and  E. Witten, {\em ``Gauge Dynamics And Compactification To Three Dimensions''}, hep-th/9607163.
\bibitem{si} K. Intriligator and N. Seiberg, Phys. Lett. B387 (1996) 513.


\bibitem{sethi} S. Sethi and  L. Susskind, Phys. Lett. B400 (1997) 265; 
T. Banks and N. Seiberg, Nucl. Phys. B497 (1997) 41, hep-th/9702187; 
N. Seiberg, {\em ``Notes on Theories with 16 Supercharges''}, hep-th/9705117.
\bibitem{DM} M.R. Douglas and G. Moore, {\em ``D-Branes, Quivers,
and ALE Instantons''},   hep-th/9603167; 
M.R. Douglas, B.R. Greene and D.R. Morrison, {\em ``Orbifold 
resolution by D-Branes''}, Nucl. Phys. B506 (1997) 84. 

\bibitem{roo} D. Z. Freedman and J. H. Schwarz, Nucl. Phys. B137 (1978) 333;
E. Bergshoeff, I. G. Koh and E. Sezgin, Phys. Lett. 155B (1985) 71; M. de Roo and P. Wagemans, Nucl. Phys. B262 (1985) 644.
 
\bibitem{fronsdal} M. Flato and C. Fronsdal, J. Math. Phys. 22 (1981) 1100;
Phys. Lett. 172B (1986) 412; Lett. Math. Phys. 2 (1978) 421; Phys. Lett. 97B (1980) 236; A. Angelopoulos, M. Flato, C. Fronsdal and D. Sternheimer, Phys. Rev.D23 (1981) 1278.  


 

 
 
  
 
\end{thebibliography}
\end{document}